# AI Detectors are Poor Western Blot Classifiers: A Study of Accuracy and Predictive Values.

Romain-Daniel Gosselin, PhD[1]

[1]Precision medicine unit, Biomedical Data Science Center (BDSC), Lausanne University Hospital (CHUV), Chemin des Roches 1a/1b CH-1010 Lausanne, Switzerland.

Tel: +4121 314 2420

Email: Romain-Daniel.Gosselin@chuv.ch.

**Sub-Title**

Western blot images created with AI cannot be confidently detected with free detectors.

**Acknowledgements**

The author would like to thank Pr. Jacques Fellay for his support.

**Conflict of interests**

The author declares no conflict of interest.

**Summary**

The recent rise of generative artificial intelligence (GenAI) capable of creating scientific images presents a challenge in the fight against academic fraud. This study evaluates the efficacy of three free web-based AI detectors in identifying AI-generated images of Western blots, which is a very common technique in biology. We tested these detectors on a collection of artificial Western blot images (n=48) that were created using ChatGPT 4 DALLE 3 and on authentic Western blots (n=48) that were sampled from articles published within four biology journals in 2015; this was before the rise of generative AI based on large language models. The results reveal that the sensitivity (0.9583 for *Is It AI*, 0.1875 for *Hive Moderation*, and 0.7083 for *Illuminarty*) and specificity (0.5417 for *Is It AI*, 0.8750 for *Hive Moderation*, and 0.4167 for *Illuminarty*) are very different. Positive predictive values (PPV) across various AI prevalence were low, for example reaching 0.1885 for *Is It AI*, 0.1429 for *Hive Moderation*, and 0.1189 for *Illuminarty* at an AI prevalence of 0.1. This highlights the difficulty in confidently determining image authenticity based on the output of a single detector. Reducing the size of Western blots from four to two lanes reduced test sensitivities and increased test specificities but did not markedly affect overall detector accuracies and also only slightly improved the PPV of one detector (*Is It AI*). These findings strongly argue against the use of free AI detectors to detect fake scientific images, and they demonstrate the urgent need for more robust detection tools that are specifically trained on scientific content such as Western blot images.

**Key words**

Research integrity; paper mills; fraud; research ethics; ROC AUC

**Introduction**

In recent years, the proliferation of artificial intelligence (AI) has introduced both unprecedented opportunities and significant challenges within the landscape of academic publishing. The emergence and fast popularisation of so-called generative AI (GenAI) such as Chat Generative Pre-trained Transformer (ChatGPT) that can generate virtually any research-relevant content – such as the text of an entire article from simple textual prompting of large language models – might help authors and editors (1). However, on the other hand, numerous dissenting voices have been raised to increase awareness about various issues linked to the use of GenAI such as authorship, plagiarism, and reliability problems (2-4). Beyond its ability to generate texts, the capacity of GenAI to produce virtually any content related to scientific research, such as images that are undetectable to the human eye, adds further anxiety about its possible fraudulent use to produce fake articles based on no existing data. In this context, one threat posed by GenAI is that it may further increase the activity of paper mills (5), which are potentially criminal for-profit companies that sell scientific manuscripts on demand and which have, disturbingly, been growing for years (6, 7). In the absence of specific models designed and trained to detect AI-generated scientific images, professionals in the publishing sector might rely on generic AI detectors that are already available on the Internet, whose efficiency at spotting scientific images is unknown.

In this study, we evaluate the performance of free web-based AI detectors in identifying AI-generated scientific images. We use the example of Western blotting, which is a technique often found in papers created by paper mills (6). It is a staple technique used in a wide range of biomedical disciplines, which is employed to detect specific proteins within a biological specimen. The output of Western blotting is an image (Western blot) that shows bands with patterns and intensities that provide qualitative and quantitative information for a target protein within specimens. Others (8) have reported, thus confirming our unpublished pilot data, that realistic Western blots can be easily created by ChatGPT.

We selected three popular detectors and used them to analyse a dataset comprising 48 AI-generated Western blot images created using ChatGPT-4 DALLE-3 and 48 genuine Western blots sourced from scientific publications in 2015 (which was years before the surge of GenAI) within four biological journals. The primary aim was to estimate the sensitivity (the proportion of AI images correctly identified as AI-generated images) and specificity (proportion of authentic images correctly identified as authentic) of these detectors, as well as their positive predictive value (PPV; proportion of positive hits that are indeed AI-generated) and negative predictive value (NPV; proportion of negative hits that are indeed authentic) across varying prevalence rates of AI-generated images. These metrics are crucial for understanding detector reliability in practical scenarios in which the prevalence of AI-generated images may vary.

Our analysis reveals an important inconsistency in performance among the three evaluated AI detectors; in particular very low PPV at realistic prevalence of AI-generated images. These results indicate that the AI-detecting tools that are currently available for free cannot be used to identify Western blot images made with GenAI in the context of peer reviewing or editorial decisions. More specific detectors that are trained on Western blot images must be urgently developed.

**Methods**

*Sample size determination.*

The number of Western blot images included in the accuracy study was determined by a sample size calculation based on the following formula (construction of an interval based on the normal approximation when the classification status is known at the time of sampling) (9):

$$n = \frac{Z^2_{(1-\alpha/2)} * S * (1-S)}{M^2}$$

where $Z$ is the standard normal value at 1-α/2, $S$ is the anticipated sensitivity (or specificity) and $M$ is the maximal margin of error. A crude pilot investigation showed that the sensitivity and specificity values of online AI detectors were relatively low when trying to identify Western blots; they detected fake images about half of the time. With $Z$=1.96, a predicted sensitivity of 0.6, and a margin of error set at 0.1, the calculation gives an estimated sample size of 92. In the absence of more precise information about the actual sensitivity and specificity of AI detectors, it was decided to evenly balance the numbers of AI-generated images and authentic Western blot images (46/46) in the final library, thus giving a prevalence of the feature to be detected (a fake image) of 0.5. We decided to extend the sample size to 48 in each group to account for potentially unusable images.

*Statistical analysis.*

Analyses were performed using R Studio (version 2023.06.2+561) or GraphPad Prism (version 10) as specified in the openly provided files. Graphics were made with GraphPad Prism. The confidence level was set at 95%, which corresponds to a false positive risk (Type I error) of 0.05 (ie. 5%). Sensitivity (the proportion of AI-generated images that will be correctly categorised), specificity (the proportion of authentic Western blots categorised as fake), and accuracy (the proportion of correctly categorised images) were calculated for each AI detector using Microsoft Excel for Mac v. 16.77.1 and confirmed using the caret R package using counts of detector outcomes as follows:

$$Sensitivity = \frac{Correctly\ categorised\ AIgenerated\ blots}{Correctly\ categorised\ AIgenerated\ blots + Falsely\ authentic\ blots}$$

$$= \frac{TP}{TP + FN}$$

$$Specificity = \frac{Correctly\ categorised\ authentic\ blots}{Correctly\ categorised\ authentic\ blots + Falsely\ AI\ categorised\ blots}$$

$$= \frac{TN}{TN + FP}$$

$$Accuracy = (0.5 * Sensitivity) + (0.5 * Specificity)$$

where TP, FN, TN, and FP indicate counts of true positive, false negative, true negative and false positive results, respectively. PPV and NPV were calculated for each AI detector on Microsoft Excel for Mac (v. 16.77.1) as follows using increasing values of AI prevalence (from 0 to 0.5):

$$PPV = \frac{Sensitivity * Prevalence}{(Sensitivity * Prevalence) + [(1 - Specificity) * (1 - Prevalence)]}$$

$$NPV = \frac{Sensitivity * (1 - Prevalence)}{[(1 - Sensitivity) * Prevalence] + [Specificity * (1 - Prevalence)]}$$

AI probabilities are given to one decimal place, whereas sensitivity, specificity, accuracy, predictive values, and receiver operator curve area under the curve (ROC AUC) are reported to four decimal places.

*Generation of fake Western blot images.*

Fake Western blot images were generated using ChatGPT 4 (https://chatgpt.com), which uses the DALLE-3 interface. The method of query was based on pilot tests that evaluated the efficacy of ChatGPT 4 at creating Western blots. We used repeated prompts asking for a "*realistic image of a Western blot*" while progressively changing the query to get new images, each time trying to guide the algorithm to a realistic Western blot image. The entire prompting history was documented and saved. Every realistic Western blot image from which four distinct bands from different lanes can be isolated was saved and used for pre-processing. Images that were not satisfactory were also saved for documentation. Four distinct chats were used to create 10 to 15 images each. The WEBP images created by ChatGPT were converted to PDF to reproduce the initial format of authentic Western blot images sampled from articles.

From a single query, the number of Western blot bands displayed on the images generated by Chat-GPT 4/DALLE-3 can greatly vary and this fickleness is not, to our knowledge, controllable by the prompt; this is apparently because ChatGPT does not easily correctly identify the terms "band" or "lane". However, the number of bands in a Western blot might influence the classification by AI detectors. Therefore, the choice was made to standardise the images by cropping them (by selective screen capturing, in PNG format) to keep only 4 lanes, with no space to the left of the left-most band or the right of the right-most band, and leaving a space equivalent to 1/10 the width of a lane above and below the bands (see Figure 1). To preserve independence between images and reduce intraclass correlation, only one cropped image was taken from each AI-generated image. To investigate whether the effect of the number of lanes in an image affected detector performance, a second data set was created by cropping these images to keep only the two lanes.

*Sampling of authentic Western blot images.*

Authentic Western blot images were sampled from published articles. To mitigate the risk of collecting AI-created images, photos were collected from articles published in 2015, which was before the rise of generative AI content that has been observed from 2020 onward. Images were sampled from four journals that frequently publish Western blot figures:

- Journal of Biological Chemistry (electronic International Standard Serial Number 1559-1182, 12 articles).
- Oncogene (electronic ISSN 1476-5594, 12 articles).
- Public Library of Open Science (PLoS) Biology (electronic ISSN 1545-7885, 12 articles).
- Cancer Research (electronic ISSN 1538-7445, 12 articles).

A systematic sampling scheme was used to collect the articles as follows. The final 2015 issue of each journal was examined, and each article with a figure containing a Western blot image was sampled (PDF file). Only one image was sampled from each article, and only blots with one single band per lane that is seen distinctly within four adjacent lanes was sampled. If figures showed the detection of multiple proteins, then priority was given to the housekeeping protein (e.g., actin, tubulin); if no housekeeping protein was displayed, the blots of the first probed protein (scanning from top to bottom) with a single band was sampled. If multiple Western blot figures were present in one article, the first Western blot appearing in the article that fulfilled the conditions above was sampled. This sampling method was applied to suitable

Western blot images from consecutive articles, chosen while moving from the beginning to the end of the issues, until the required number of images had been obtained.

The following exclusion criteria were predefined and applied:

- Blots from immunoprecipitation or pull-down assays (because they might have specific background or signal intensity).
- Blots show in colour of with white bands against a dark background (these are natural luminescent image acquisition, but these are colour-reversed images compared to convention).
- Images with less than 4 lanes.
- Images with inserts such as a highlighted area, framed zones, arrows, or text.
- Conference abstracts, reviews, or perspectives articles.

Authentic Western blot images were captured using selective screenshots, and the same protocol was employed with AI-generated blots (i.e., keeping only 4 lanes, leaving no space left and right and a space corresponding to 1/10 of a lane width above and below the bands).

*Selection and performance analysis of AI detectors on Western blots displaying 4 or 2 lanes.*
A Google search using the query "Free AI image detector" was performed on May 17$^{th}$ 2024 in Lausanne (Switzerland). The first 3 results that corresponded to free websites that did not require a subscription were used: *Is It AI?* (https://isitai.com/ai-image-detector/), *Hive Moderation* (https://hivemoderation.com/ai-generated-content-detection), and *Illuminarty* (https://app.illuminarty.ai/). Each image was scanned with all three of the AI detectors. The output of the AI detectors is a probability, given as a percentage, that the image is AI-generated. Images were classified as a true positive, true negative, false positive, or false negative, with positivity being defined as a detector output above 50% of probability. In addition, predictive values (PPV and NPV) were also calculated for each detector.

Two detector outputs were included in the analysis: 1) image classification determined by the detector outputs, which were used in the confusion matrix to calculate the sensitivity, specificity, accuracy, and predictive values; and 2) the absolute AI probability returned by the detectors.

*Data storage and sharing*

AI images were saved in the WEBP format generated by ChatGPT, then converted to PDF format, and then cropped screenshot images in PNG format were created and stored for analysis. Authentic Western blot images were saved, processed, stored, and analysed in PNG format from screenshots; no digital modifications, such as contrast, were applied to the images. All data were saved, stored, and shared on a publicly available Figshare repository (https://figshare.com) as follows:

A data folder accessible at https://doi.org/10.6084/m9.figshare.26300464 containing:

- The entire prompting history used to create images.
- A spreadsheet (Excel) that summarises all quantitative analyses.
- A spreadsheet (Excel) that provides details of all sampled articles.
- Comma-separated values (CSV) files for each specific data set.
- The R codes used to analyse the data.
- GraphPad Prism files used to generate the figures.

A data folder accessible at https://doi.org/10.6084/m9.figshare.26300515 containing:

- The cropped versions of authentic Western blots.
- The full AI-generated images and their cropped versions.

- The unused (failed) ChatGPT 4 images.

**Results**

*AI detectors showed highly varied abilities to distinguish between artificial and authentic Western blots.*

Quantification of the AI probabilities that were returned by AI detector following the assessment of Western blots with four lanes showed that AI probability values spanned a very wide range for each AI detector (Figure 2A). *Illuminarty* gave the highest average AI probabilities (median=86.2, IQR [42.2–98.7] for AI Western blots, median=81.1, IQR [19.2–99.1] for authentic Western blots). The lowest AI probabilities were returned by *Hive Moderation* (median=17.0, IQR [9.5–30.3] for AI Western blots, median=18.2, IQR [8.3–37.0] for authentic Western blots). *Is It AI* produced an intermediate output that was visibly dissimilar between AI Western blots (median=85.1, IQR [69.8–93.1]) and authentic Western blots (median=47.3, IQR [26.9–66.0] for authentic Western blots). This variability in detector output was reflected in the formal confusion matrices and accuracy analyses presented in Figure 3B. The proportions of false positives were 22/48 for *Is It AI*, 6/48 for *Hive Moderation*, and 28/48 for *Illuminarty*, and the proportions of false negatives were 2/48 for *Is It AI*, 39/48 for *Hive Moderation*, and 14/48 for *Illuminarty*. As presented in Figure 4A, the consequence of these high rates of misclassifications is that detector performances fell often short of usual standards (*Is It AI*: sensitivity=0.9583, specificity=0.5417, accuracy=0.7500, ROC AUC=0.9028, 95% CI [0.8435, 0.9621]; *Hive Moderation*: sensitivity=0.1875, specificity=0.8750, accuracy=0.5313, ROC AUC=0.5100, 95% CI [0.3930, 0.6270]; *Illuminarty*: sensitivity=0.7083, specificity=0.4167, accuracy=0.5625, ROC AUC=0.5449, 95% CI [0.4276, 0.6622]).

As shown in Figure 2B, when sizes of the scanned Western blots were cropped from 4 to 2 lanes, the AI probabilities of all detectors were markedly reduced (*Is It AI*: median=62.8, IQR [38.5–80.8] for AI Western blots, median=11.7, IQR [6.2–28.8] for authentic Western blots; (*Hive Moderation*: median=6.2, IQR [3.1–18.0] for AI Western blots, median=12.2, IQR [5.6–24.1] for authentic Western blots); *Illuminarty*: median=15.4, IQR [7.8–43.5]) for AI Western blots, median=9.1, IQR [2.4–29.0] for authentic Western blots). Consequently, there was a dramatic increase the number of false negative results delivered (*Is It AI*: 20/48; *Hive Moderation*: 46/48; *Illuminarty*: 38/48) and a reduction in the number of false positive results (*Is It AI*: 4/48, *Hive Moderation*: 4/48, *Illuminarty*: 8/48) (Figure 3C). Figure 4B shows that although the overall test accuracies were not noticeably impacted by reducing the number of bands (*Is It AI*: accuracy=0.7500, ROC AUC=0.8974, 95% CI [0.8361, 0.9586]; *Hive Moderation*: accuracy=0.4792, ROC AUC=0.6252, 95% CI [0.5118, 0.7386]; *Illuminarty*: accuracy=0.5208, ROC AUC=0.6248, 95% CI [0.5115, 0.7380]), reduced sensitivity was observed (*Is It AI*: 0.5833; *Hive Moderation*: 0.0417; *Illuminarty*: 0.2083) along with an increased specificity (*Is It AI*: 0.9167; *Hive Moderation*: 0.9167; *Illuminarty*: 0.8333).

*Study of positive and negative predictive values of AI detectors.*

Sensitivity and specificity are intrinsic properties of the detectors that describe their ability to correctly classify images as authentic or fake. However, in editorial contexts where the question is whether a given image can be deemed authentic, PPV and NPV would be more useful; these measures are functions of both sensitivity and specificity, but also of the prevalence of AI images in the literature. Therefore, we calculated the PPV and NPV for each of the three AI detectors in scenarios in which AI prevalence sequentially increased from 0 up to 0.5 (Figure 5). Data from Western blots with 4 lanes (Figure 5A) showed that the PPVs of the three detectors were very low when the AI prevalence was set below 0.1, with a maximum value observed for *Is It AI* at 0.1885 when AI prevalence was set at 0.1. This indicates that most of the Western blots most categorised as AI-generated would be false positives. When reducing further AI prevalence in the simulations, PPV become dramatically low. For example, at a prevalence of 0.005 (meaning that we expect 1 Western blot out of 200 to be AI-generated), PPVs were 0.0104 for *Is It AI*, 0.0075 for *Hive Moderation* and 0.0061 for *Illuminarty*. Conversely, NPV was higher than 0.9 for all detectors at this realistic AI prevalence, indicating that false negatives would be rare in this scenario. At higher values of AI prevalence, the PPVs of all detectors increased steadily as NPV decreased, although a discrepancy was observed between the relatively high NPV of *Is It AI* (NPV=0.9285 at AI probability=0.5) and the NPVs of *Hive Moderation* (NPV=0.5185 at AI probability=0.5) and *Illuminarty* (NPV=0.5185 at AI probability=0.5). PPVs and NPVs calculated using Western blots with two lanes (Figure 5B) showed patterns for *Hive Moderation* and *Illuminarty* that were similar to those obtained from four-lane blots. However, *Is It AI* shows a higher PPV than those of the two other detectors, even for low values of AI prevalence, and a lower NPV than when blots with four lanes were analysed.

**Discussion**

The present study addresses a critical issue about the efficacy of current free AI detectors in recognising AI-based image forgery in scientific publications. Using Western blots as an example, we demonstrate that such AI detectors are inadequate for aiding editorial vigilance in detecting images created by GenAI. It remains uncertain whether AI-generated Western blots have already infiltrated the biological literature, and the current detection tools would not yet be capable of efficiently identifying them. Our findings that detectors show different performances and often misclassify Western blot images largely align with previous studies on AI-generated texts (10-15) and likely extend to other types of non-textual scientific content.

One novel finding is the large variation in sensitivity and specificity values that was observed across the three detectors tested. AI detectors often generated completely opposing AI probability outputs for one given image. This emphasises the importance of the different mathematical architectures on which the various tools are based. Beyond their inconsistency, the values obtained for these metrics often fell well below the acceptable thresholds for reliable detection, indicating frequent image misclassification. This result is in clear contrast with the high sensitivity and specificity reported for the detection of text created by AI (16, 17), demonstrating the relative immaturity of AI image detectors. When examining PPV and NPV, which would be the essential measures for editorial decision-making, the results indicated that the PPV was low at realistic probabilities of AI images in the literature. This suggests that concluding of falseness of a Western blot image based on a high AI probability given by AI detectors would often be misleading. The actual prevalence of AI-generated images in published articles is currently unknown, and it is challenging to estimate in the absence of reliable detectors. Nevertheless, the existing estimates indicate that the prevalence of text created by AI in publications is already exceeding 10–30% (12, 17-19), and the rate of inappropriate image duplication is projected to lie around 5% (20). These figures suggest that AI-generated images might already be well entrenched in the scientific literature. Notably, our results also indicate that reducing the number of lanes in the images resulted in decreased detector sensitivity and increased specificity. This relationship between Western blot complexity in terms of band richness and detector output will have to be accounted for when integrating AI detection in editorial process. The fact that image classification may be affected by image editing further highlights the critical importance of thoroughly documenting and providing data for all steps of image pre-processing, as well as the importance of systematically sharing unprocessed raw images.

One way to help improve the detection process could be to use automated AI detection as a subsequent step following a human-led preliminary identification process that is applied only to suspicious articles. This approach mirrors the traditional distinction in epidemiology between large-scale screening tests that are performed agnostically versus diagnostic tests performed on symptomatic patients. Automated AI detection would be applied exclusively to articles flagged with potential AI indicators, such as authors with multiple article retractions or research fields that are known for frequent publication of AI-generated images. This two-step strategy could enhance the PPV by increasing the probability of AI in articles in the sample. The improvement of AI detection will also fundamentally hinge on the inclusion of Western blot images in training sets used to develop image detectors. Improving the performance of detector algorithms on scientific images, such as Western blots, requires their rigorous design and training on large collections of authentic Western blot images.

The first limitation of this study pertains to the potential non-representativeness of the three detectors included. It is imaginable that other detectors, particularly those that require a subscription, might exhibit superior performance (15). In relation to this first limitation, we have restricted our sampling images to Western blots created by ChatGPT due to the dominant popularity of this software, but images from other generators might be classified differently by AI detectors. Secondly, images were classified as AI generated if their AI probability was at least 50%. This threshold is expected to significantly impact the rates of false positives and false negatives, consequently affecting the calculation of performance metrics (12). Future studies should explore the impact of varying thresholds on the ability of detectors to distinguish between authentic and fake images. We also used PNG images from screenshots, not only because we had no access to raw images but also because we assumed that a similar approach would be used by peer-reviewers and editors. However, image format (e.g., JPEG, TIFF) might influence the detector output. Future detector algorithms should therefore be trained using various formats. Finally, we had no indication about whether the authentic Western blots included during our sampling had been acquired using photographic film exposure or using a charge-coupled device (CCD) camera. It is possible that images obtained through different acquisition methods would give different AI probabilities when scanned by AI detectors.

In conclusion, this study uses the example of Western blots to raise awareness about the urgent need for more effective AI detectors that are specifically designed and trained to reveal fake scientific images. The implications of the current findings are profound for editors, publishers, and reviewers tasked with maintaining the integrity of the scientific literature. Enhanced AI detection capabilities, coupled with rigorous editorial policies and reviewer training, are vital for upholding the standards of scientific publishing.

**Figure Legend:**

**Figure 1: General design of the study.** Western blots generated by AI (left, N=48) were created using ChatGPT 4 and the downloaded images (WEBP format) were converted to PDF before being cropped by selective screen capturing to keep only individual bands on four lanes and saved as a PNG. Authentic Western blot images (right, N=48) were sampled from downloaded articles in articles published in 2025 in four journals. Images were obtained by selective screen capturing of individual bands on four lanes. Both AI and authentic images of Western blots with two lanes were obtained by cropping (selective screen capturing, red insert)

the four-lane blots. All images were scanned using three online AI detectors. The AI probability obtained for each image was both reported and used to classify them in the confusion matrix as true or false positives or negatives, and to calculate detector performances.

**Figure 2: Distribution of AI probability returned by AI detectors.** The violin plots show the density distribution of AI probabilities (given as percentage on y-axis), with individual data points shown (each dot represents a single Western blot image). For each AI detector (*Is It AI* in blue, *Hive Moderation* in green, and *Illuminarty* in red), the group of AI-generated Western blots is shown on the left (N=48) and the authentic Western blots is shown on the right (N=48). Panel A shows 4-lane Western blots, panel B shows 2-lane Western blots. The horizontal dotted line indicates the limit (50% probability) set to define positive results.

**Figure 3: Confusion matrices.** The confusion matrices show the quality of the classification systems. For each table, the genuine status of the images is given on the left-hand side and the status given by the detector is shown on the top. The four possible outcomes of the 2x2 matrices are shown in A. TP=true positives, FP=false positives, FN=false negatives, TN=true negatives. The counts of different outcomes are shown for Western blots with four lanes (B) and two lanes (C).

**Figure 4: Performance and receiver operator curves (ROC) of AI detectors.** The tables on the left show the sensitivity, specificity, and accuracy of each AI detector. On the right, ROC curves are given for each detector (*Is It AI* in blue, *Hive Moderation* in green, *Illuminarty* in red). The area under the curve (AUC) is given as a point estimate with 95% confidence interval between brackets. Matrices and ROC in A were obtained from Western blots with four lanes, those in B were obtained from Western blots with two lanes.

**Figure 5: Positive and negative predictive values.** The graphs show the positive predictive value (PPV, left) and negative predictive value (NPV, right) calculated for different theoretical prevalences of AI-generated images between 0 and 0.5. Graphs show predictive values obtained from Western blots with four lanes (A) or two lanes (B) for the three AI detectors (*Is It AI* in blue, *Hive Moderation* in green, *Illuminarty* in red).

**Figure 1**

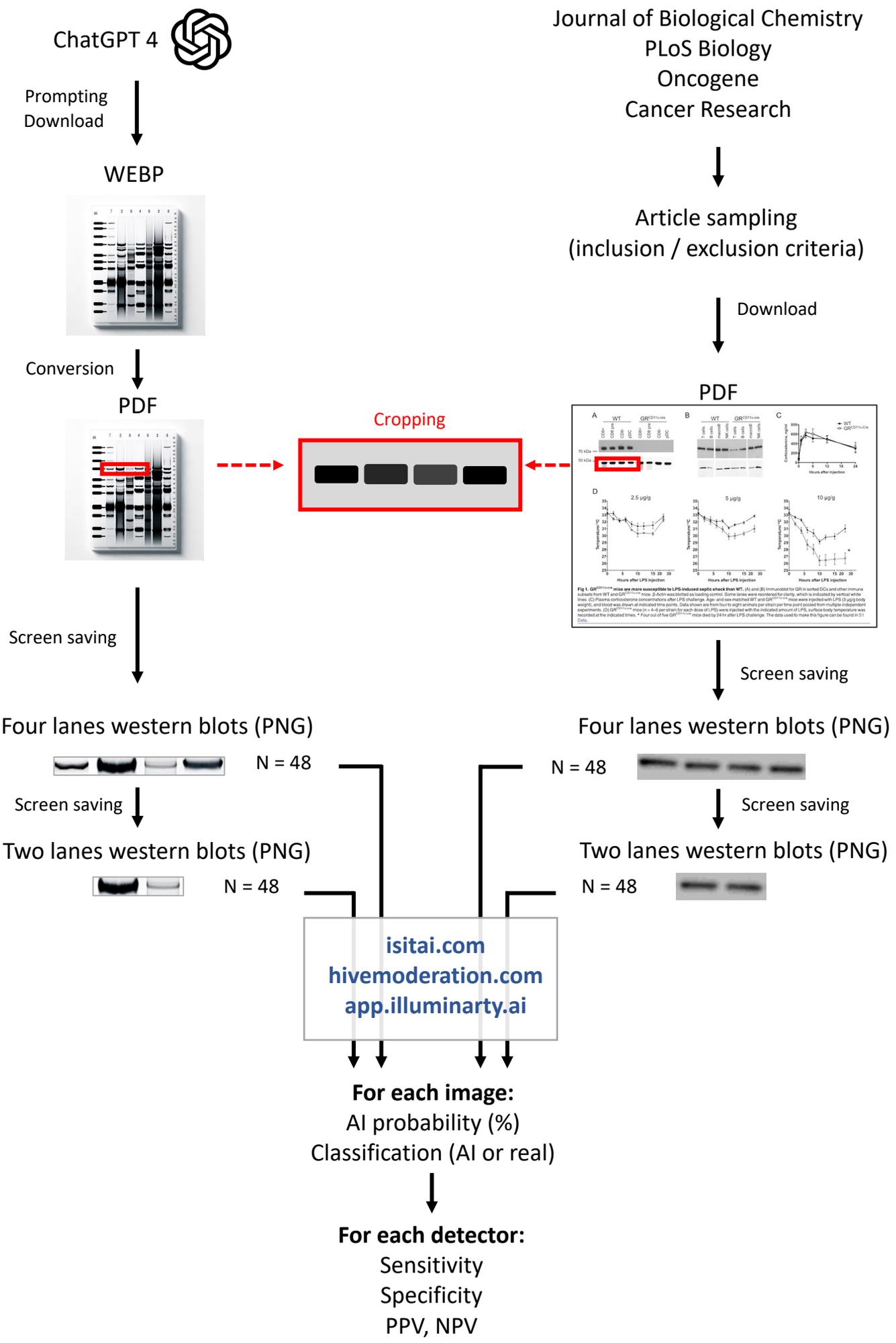

**Figure 2**

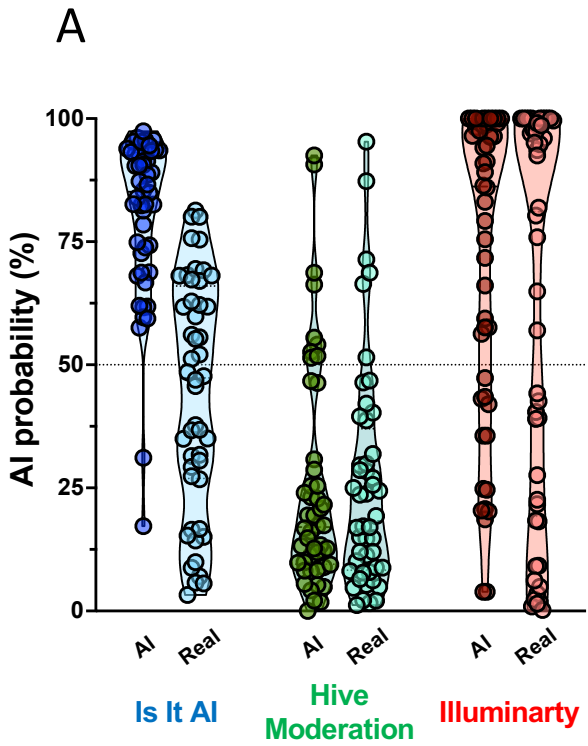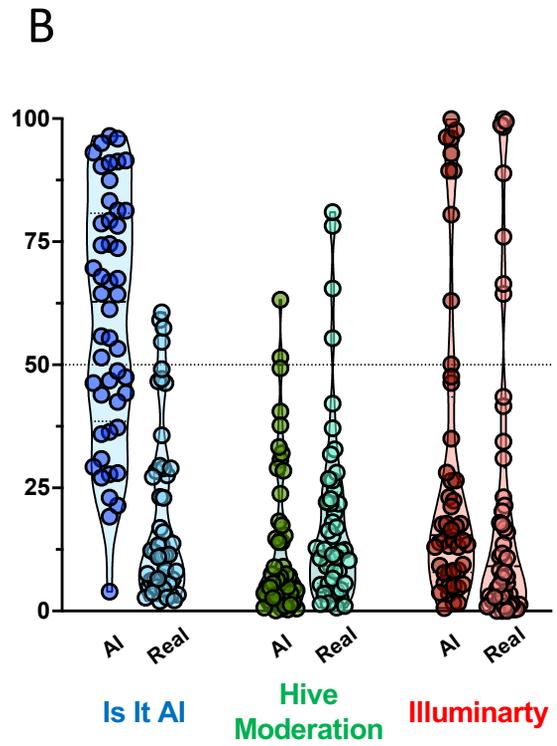

**Figure 3**

**A**

|  | Detector output | |
|---|---|---|
|  | AI | Real |
| Actual status — AI | TP | FN |
| Actual status — Real | FP | TN |

**B** **Four lanes western blots**

Is It AI

|  | Detector output | |
|---|---|---|
|  | AI | Real |
| Actual status — AI | 46 | 2 |
| Actual status — Real | 22 | 26 |

Hive Moderation

|  | Detector output | |
|---|---|---|
|  | AI | Real |
| Actual status — AI | 9 | 39 |
| Actual status — Real | 6 | 42 |

Illuminarty

|  | Detector output | |
|---|---|---|
|  | AI | Real |
| Actual status — AI | 34 | 14 |
| Actual status — Real | 38 | 20 |

**C** **Two lanes western blots**

Is It AI

|  | Detector output | |
|---|---|---|
|  | AI | Real |
| Actual status — AI | 28 | 20 |
| Actual status — Real | 4 | 44 |

Hive Moderation

|  | Detector output | |
|---|---|---|
|  | AI | Real |
| Actual status — AI | 2 | 46 |
| Actual status — Real | 4 | 44 |

Illuminarty

|  | Detector output | |
|---|---|---|
|  | AI | Real |
| Actual status — AI | 10 | 38 |
| Actual status — Real | 8 | 40 |

# Figure 4

## A

### Four lanes western blots

| | Is It AI | Hive Moderation | Illuminarty |
|---|---|---|---|
| Sensitivity | 0.9583 | 0.1875 | 0.7083 |
| Specificity | 0.5417 | 0.8750 | 0.4167 |
| Accuracy | 0.7500 | 0.5313 | 0.5625 |

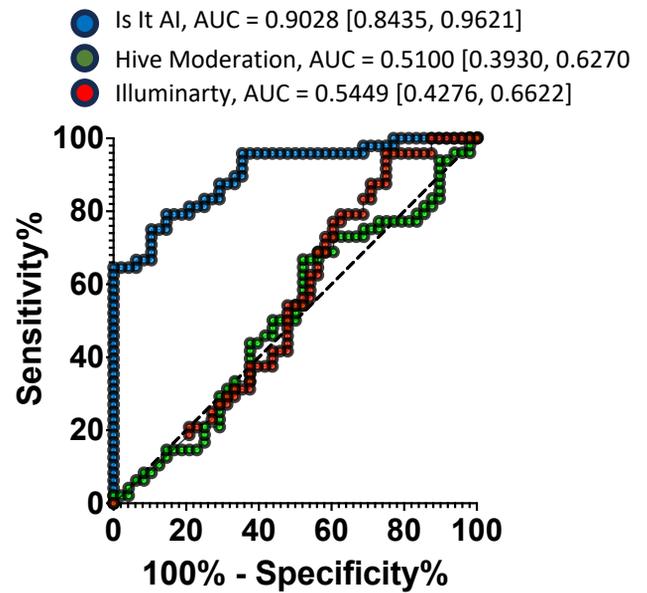

## B

### Two lanes western blots

| | Is It AI | Hive Moderation | Illuminarty |
|---|---|---|---|
| Sensitivity | 0.5833 | 0.0417 | 0.2083 |
| Specificity | 0.9167 | 0.9167 | 0.8333 |
| Accuracy | 0.7500 | 0.4792 | 0.5208 |

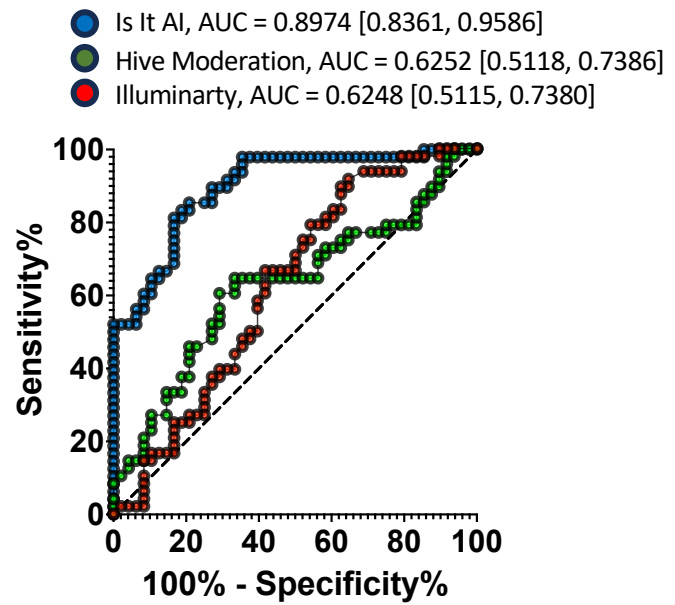

**Figure 5**

A

## Four lanes western blots

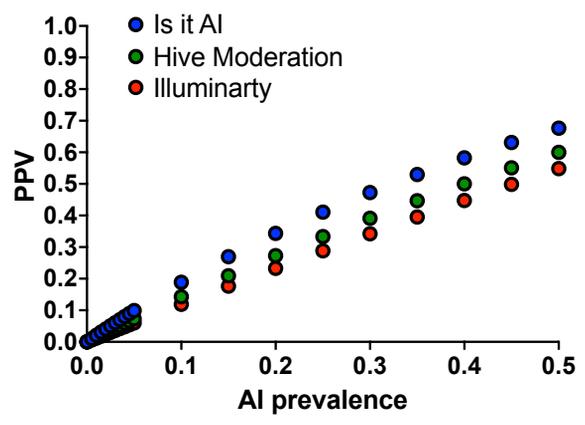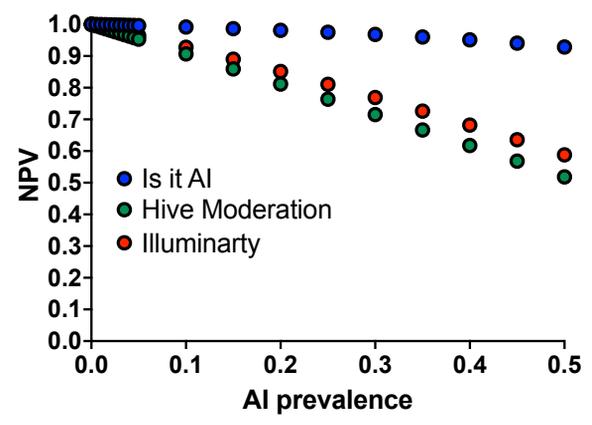

B

## Two lanes western blots

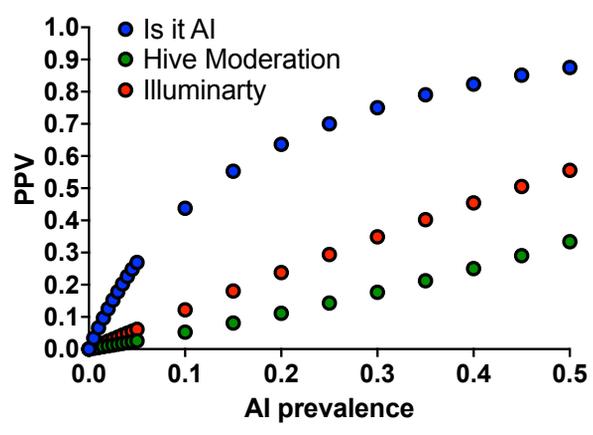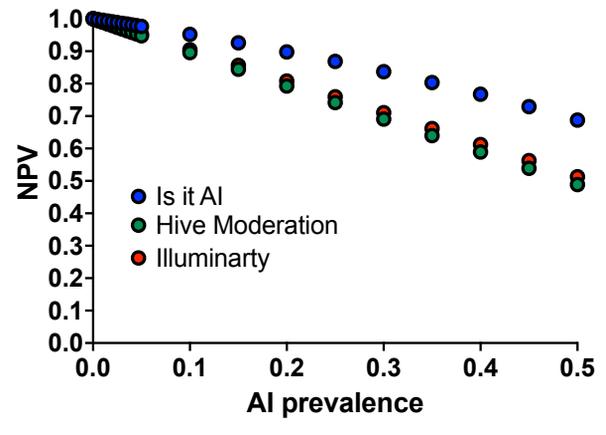